\documentclass[twocolumn,epsf,epsfig,showpacs,prl]{revtex4}
\usepackage{graphicx}
\usepackage{dcolumn}
\usepackage{bm}

\begin{document}

\title{Lorentz Violation for Photons and Ultra-High Energy Cosmic Rays}
\author{Matteo Galaverni$^{a,b,c}$, G{\"u}nter Sigl$^{d,e}$}

\affiliation{$^a$INAF-IASF Bologna, 
via Gobetti 101, I-40129 Bologna - 
Italy}
\affiliation{$^b$Dipartimento di Fisica, Universit\`a di Ferrara,
via Saragat 1, I-44100 Ferrara - Italy}
\affiliation{$^c$INFN, Sezione di Bologna,
Via Irnerio 46, I-40126 Bologna - Italy}
\affiliation{$^d$II. Institut f\"ur theoretische Physik, Universit\"at Hamburg,
Luruper Chaussee 149, D-22761 Hamburg, Germany}
\affiliation{$^e$
APC~\footnote{UMR 7164 (CNRS, Universit\'e Paris 7,
CEA, Observatoire de Paris)} (AstroParticules et Cosmologie),
10, rue Alice Domon et L\'eonie Duquet, 75205 Paris Cedex 13, France}

\begin{abstract}
Lorentz symmetry breaking at very high energies may lead
to photon dispersion relations of the form 
$\omega^2=k^2+\xi_n k^2(k/M_{\rm Pl})^n$
with new terms suppressed by a power $n$ of the Planck mass $M_{\rm Pl}$. We show
that first and second order terms of size 
$\left|\xi_1\right|\gtrsim10^{-14}$ and $\xi_2\lesssim-10^{-6}$, 
respectively, would lead to a photon component in cosmic rays above
$10^{19}\,$eV that should already have been detected, if corresponding
terms for $e^\pm$ are significantly smaller.
This suggests that LI breaking suppressed up to second order
in the Planck scale are unlikely to be phenomenologically viable
for photons.
\end{abstract}


\maketitle

{\it Introduction.} 
Many Quantum Gravity theories suggest the breaking of Lorentz invariance
(LI) with the strength of the effects increasing with energy. The most promising
experimental tests of such theories, therefore, exploit the highest energies at
our disposal which are usually achieved in violent astrophysical processes. 
If LI is broken in form of non-standard dispersion
relations for various particles, absorption and energy loss
processes for high energy cosmic radiation can be modified~\cite{Coleman:1998ti}.
Conversely, experimental confirmation that such processes occur at the
expected thresholds would allow to put strong constraints on such LI
breaking effects. This was shown in case of ultra-high-energy cosmic rays
producing pions by the Greisen-Zatsepin-Kuzmin (GZK) effect~\cite{gzk} above the
threshold at $\sim 7\times 10^{19}$ eV and in case of pair production of high energy 
photons with the diffuse low energy photon background~\cite{Aloisio:2000cm}.

While the thresholds of electron-positron pair production by high energy $\gamma-$rays
on low energy background photons have not yet been experimentally confirmed
beyond doubt, constraints on LI breaking for photons have been
established based on the very existence of TeV $\gamma-$rays from astrophysical
objects~\cite{Jacobson:2002hd}.

Here we exploit the fact that if pair production of high energy $\gamma-$rays
on the cosmic microwave background (CMB) would be inhibited above $\sim10^{19}\,$eV,
one would expect a large fraction of $\gamma-$rays in the cosmic ray flux at these energies, independent on where the real pair production threshold
is located. Based on the fact that no significant $\gamma-$ray fraction is
observed, we derive limits on LI violating parameters for photons that
are more stringent than former limits. These limits do not depend on the poorly
known strength of the astrophysical radio background.

Hybrid detectors begin to put constraints on the composition of 
cosmic rays at highest energies. Particularly it is already possible 
to put upper limits on the fraction of photons on the $10\%$ level at 
energies above $10^{19}$ eV using Auger hybrid observations~\cite{Abraham:2006ar},
AGASA~\cite{Shinozaki:2002ve,Risse:2005jr,Rubtsov:2006tt} and
Yakutsk~\cite{Rubtsov:2006tt,Glushkov:2007ss} data. Above $10^{20}\,$eV,
the current upper limit is $\sim40$\%~\cite{Rubtsov:2006tt}. In fact,
the latest upper limits from the surface detector data of the Pierre Auger
observatory are already at the level of $\sim2$\% above
$10^{19}$ eV~\cite{Healy:2007ef}.
In the next few year these constraints will improve
with statistics: The Pierre Auger experiment can reach a sensitivity of
$\sim0.3$\% within a few years and $\sim0.03$\% within 20 years around
$10^{19}\,$eV, and a sensitivity at the $10\%$ level around
$10^{20}$ eV within 20 years~\cite{Risse:2007sd}.

Neutral pions created by the GZK effect decay into ultra-high
energy photons. They subsequently interact
with low-energy background photons of the CMB and the universal radio
background (URB) through pair production,
$\gamma\,\gamma\rightarrow e^+\,e^-$. This leads to the development 
of an electromagnetic cascade and suppresses the photon flux above the
pair production threshold on the CMB of $\sim10^{15}\,$eV. Above $\sim10^{19}\,$eV
the interaction length for photons is smaller than a few Mpc, whereas for 
nucleons above the GZK threshold at $\sim7\times10^{19}\,$eV it is of the order
of 20 Mpc. As a result, the photon fraction theoretically expected is
smaller than $\sim1$\% around $10^{19}\,$eV, and smaller than $\sim10$\%
around $10^{20}\,$eV~\cite{Sigl:2007ea,Gelmini:2007jy}, in agreement with
experimental upper limits.

\begin{figure}
\includegraphics[scale=0.5]{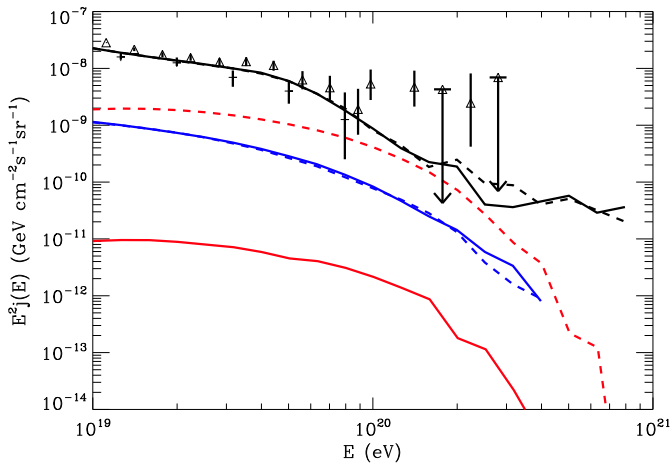}
\caption{Fluxes of protons (black), photons (red)
   and neutrinos per flavor (blue) for uniform $E^{-2.6}$
   proton injection between $10^{19}$ and $10^{21}$ eV up to
   redshift 3. AGASA data~\cite{Shinozaki:2006kk} are shown as
   triangles, HiRes data~\cite{Abbasi:2002ta} as crosses. Solid:
   with CMB and the minimal version of the
   universal radio background, based on observations~\cite{radio-obs};
   dashed: without any pair production by photons above $10^{19}\,$eV.} 
\label{fig1}
\end{figure}

\begin{figure}
\includegraphics[scale=0.5]{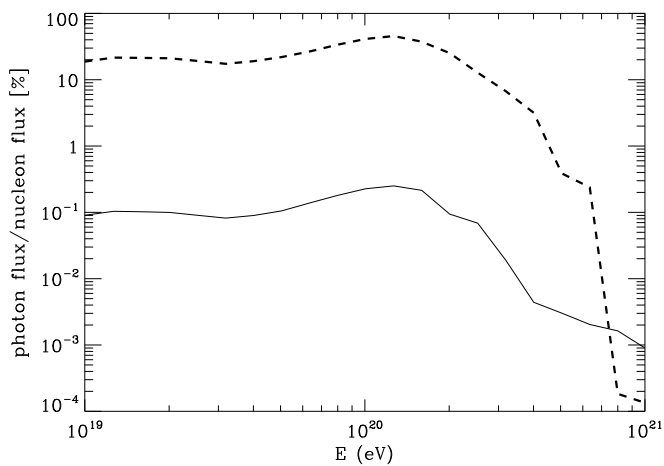}
\caption{The ratio of the integral photon to primary cosmic ray flux
  above a given energy as a function of that energy for the
   two scenarios shown in Fig.~\ref{fig1}.} 
\label{fig2}
\end{figure}

The breaking of Lorentz invariance, by modifying the dispersion relation
for photons, would affect the energy threshold for pair production.
If the change in the dispersion relation is sufficiently large, pair production 
can become kinematically forbidden at ultra-high energies and such photons
could reach us from cosmological distances. As a consequence, at
least if ultra-high energy cosmic rays consist of mostly protons,
one would expect
a significant photon fraction in cosmic rays above $10^{19}\,$eV, in conflict with
experimental upper limits. Figs.~\ref{fig1} and~\ref{fig2} which were obtained
with the CRPropa code~\cite{crpropa,Armengaud:2006fx}
show that the ratio of the integral
photon to primary cosmic ray flux above $10^{19}\,$eV would be $\simeq20$\%,
and thus higher than the above mentioned experimental upper limits. In this
scenario, we have used a relatively steep proton injection spectrum
$\propto E^{-2.6}$. Harder injection spectra also give acceptable fits
above $\sim10^{19}\,$eV, as well as higher photon fractions due to
increased pion production~\cite{Gelmini:2007jy}. Whereas for pair production without LI breaking, the predicted photon fraction always stays below experimental
upper limits, harder injection spectra and larger maximal energies
in the absence of pair production would overshoot the experimental limits
even more than in Figs.~\ref{fig1} and~\ref{fig2}.

Therefore, LI violating parameters for photons are
constrained by the requirement that pair production be allowed
between low energy background photons and photons of energies between
$10^{19}$ eV and $10^{20}$ eV.
We will assume that pion production itself is
not significantly modified and that the modifications of the
dispersion relations of electrons and positrons are significantly
smaller than for photons. This is consistent since the photon content
of other particles is on the percent level~\cite{Gagnon:2004xh}.
 
{\it Formalism.}
We denote the 4-momenta with $(\omega,\bf{k})$ for the  ultra-high-energy photon, $(\omega_b,\bf{k_b})$ for the background 
photon, and $(E_\pm,\bf{p}_\pm)$ for the electron and positron, respectively.

We consider the following modified dispersion relations for photons, electrons
and positrons:
\begin{eqnarray}
\label{photonDR}
\omega^2&=&k^2 + \xi_n k^2 \left(\frac{k}{M_{\rm pl}}\right)^n\,,\nonumber\\
E_\pm^2&=&p_\pm^2 + m_e^2+\eta_n^\pm p_\pm^2 \left(\frac{p_\pm}{M_{\rm pl}}\right)^n\,,
\end{eqnarray} 
with $n\geq 1$ and where $M_{\rm pl}\simeq10^{19}$ GeV and $m_e$ are the Planck
mass and the electron mass, respectively.

Using the exact relation for energy-momentum conservation, 
the kinematic relation for the decay of a neutral pion of mass $m_\pi$
into two $\gamma-$rays of energy-momentum $(\omega_1,{\bf k}_1)$
and $(\omega_1,{\bf k}_2)$, respectively, and equal helicity is 
$2\omega_1\omega_2-2{\bf k}_1\cdot{\bf k}_2+\xi_n(k_1^{n+2}+k_2^{n+2})
/M_{\rm Pl}^n=m_\pi^2$. For $\left|\xi_n\right|\lesssim1$, the absolute values of
the LI violating terms are always much smaller than the ones of
$\omega_1\omega_2$ and ${\bf k}_1\cdot{\bf k}_2$,
which themselves are much larger than $m_\pi^2$ in most of the
phase space. Therefore, the kinematics of pion decay is not significantly modified.

This is different for pair production by photons: 
Exact energy momentum conservation implies that 
$\left(\omega+\omega_b\right)^2-\left(\bf{k}+\bf{k_b}\right)^2 
=\left(E_++E_+\right)^2-\left(\bf{p_+}+\bf{p_-}\right)^2$. The
left hand side is maximized for anti-parallel initial photon momenta
(head-on collision) and the right hand side is minimized for
parallel final momenta of the pair~\cite{Jacobson:2002hd,Mattingly:2002ba}.
Writing $p_+=yk$, $p_-=(1-y)k$ with $0\leq y\leq1$, 
assuming relativistic leptons and using
$\omega\gg\omega_b$, after some algebra we thus obtain at the threshold
\begin{equation}\label{eq_thresh}
  \kappa_n\,k^2
  \left(\frac{k}{M_{\rm Pl}}\right)^n+4k\omega_b-\frac{m_e^2}{y(1-y)}=0\,,
\end{equation}
where
\begin{equation}
  \kappa_n\equiv\xi_n-\eta_n^+y^{n+1}-\eta_n^-(1-y)^{n+1}
\end{equation}
and the asymmetry $y$ in the final momenta at threshold is determined
by maximizing the left hand side of Eq.~(\ref{eq_thresh}).
For example, if $\eta_n^+=\eta_n^->-2^{n+3}
\left(m_e/k\right)^2\left(M_{\rm pl}/k\right)^n/n(n+1)$, then $y=\frac{1}{2}$.

Introducing $x\equiv k/k_0$ with $k_0\equiv m_e^2/\left[4y(1-y)\omega_b\right]$,
Eq.~(\ref{eq_thresh}) can be rewritten as
\begin{equation}
\label{xEQ1}
 \alpha_{n} x^{n+2}+x-1=0\,,
\end{equation}
where
\begin{equation}\label{alpha_def}
\alpha_n\equiv\kappa_n\frac{k_0}{4\omega_b}\left(\frac{k_0}{M_{\rm pl}}\right)^n\,.
\end{equation}

If $\xi_n=\eta_n^+=\eta_n^-=0$ we have $\alpha_n=0$ and $y=\frac{1}{2}$
and thus the usual threshold for pair production in Lorentz invariant theory,
$k=m_e^2/\omega_b$. Furthermore, if the LI violating terms in the electron
and positron dispersion relations are smaller than the photon terms,
$|\eta_n^\pm|\lesssim\xi_n$, then $\kappa_n\simeq\xi_n$ and we
will obtain constraints essentially on the photon terms $\xi_n$. If not
otherwise stated we will make this assumption in the following.

If $\alpha_n>0$, Eq.~(\ref{xEQ1})
admits one real positive solution $x_n^l(\alpha_n)<1$ for each value of $\alpha_n>0$.
Therefore, for photons with a {\em positive} LI violating term 
in the modified dispersion relation Eq.~(\ref{photonDR}), pair production
is kinematically allowed above a threshold $k_0x_n^l(\alpha_n)<k_0$.

\begin{figure}
\includegraphics[scale=1]{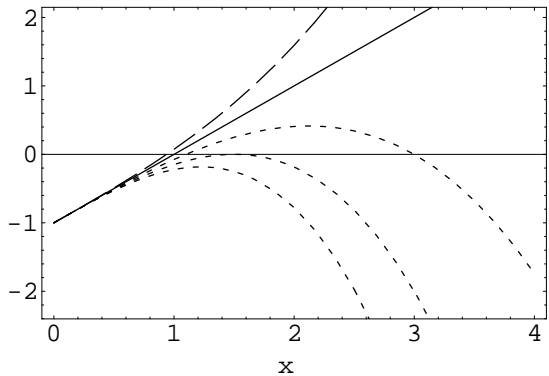}
\caption{The left hand side of Eq.~(\ref{xEQ1}) for various cases
for $n=1$. Red: photons with a positive LI breaking term $\alpha_1=2/27$;
Black: photons with unbroken LI, $\alpha_1=0$; Blue: photons with a
negative LI breaking term, with $\alpha_1=-6/27$, $-$4/27, $-$2/27,
in ascending order. Pair production is kinematically allowed for values
of $x\equiv k/k_0$ for which the curves are positive.} 
\label{fig3}
\end{figure}

Otherwise, if the coefficient of $x^{n+2}$ in Eq.~(\ref{xEQ1}) is negative,
this equation has real solutions only if
$\left|\alpha_n\right|\leq \alpha_n^{\rm cr}\equiv
\left(n+1\right)^{n+1}/\left(n+2\right)^{n+2}$.
In particular, if $\left|\alpha_n\right|=\alpha_n^{cr}$ there is only one real solution
and pair production is kinematically allowed only for a particular value of 
the momentum of the ultra-high-energy photon. If $\left|\alpha_n\right|<\alpha_n^{cr}$,
there are two real solutions, $0<x_n^l\left(\alpha_n\right)<x_n^u\left(\alpha_n\right)$,
and thus pair production is only allowed in the range of energies
$k_0x_n^l\left(\alpha_n\right)\leq\omega\leq k_0x_n^u\left(\alpha_n\right)$.
These two cases are summarized in Fig.~\ref{fig3}.

Requiring pair production to be allowed, we obtain constraints only from
photons with a negative sign in 
the modified dispersion relation, because for photons with a positive 
LI breaking term, pair production is allowed for any value of $\alpha_n$
above $k_0x_n^l(\alpha_n)<k_0$.
We also stress that photons with negative LI breaking term
are stable against {\em photon decay} ($\gamma\rightarrow e^+\,e^-$)
and  {\em photon splitting} ($\gamma\rightarrow N\,\gamma$).

Requiring the interaction of ultra-high-energy photons,
$10^{19}\,\mbox{eV}\lesssim k\lesssim 10^{20}\,\mbox{eV}$, with CMB photons
of energy $\omega_b\simeq 6\times 10^{-4}$ eV corresponds to requiring
that pair production is kinematically allowed for
$2.3\times 10^{4}\lesssim x\lesssim 2.3\times 10^{5}$.
Since photons with a negative LI breaking term in the dispersion relation 
have both a lower and an upper energy threshold for pair production,
denoted by $x_n^l(\alpha_n)$ and  $x_n^u(\alpha_n)$, respectively,
we have the two conditions $x_n^l(\alpha_n)\lesssim2.3\times 10^{4}$
and $2.3\times10^{5}\lesssim x_n^u(\alpha_n)$. These will lead to
constraints on $\alpha_n$ and thus $\xi_n$.

{\it Constraints on Lorentz invariance breaking to first order in the Planck mass.}
In this case $n=1$ and the first condition,
$x_1^l\left(\alpha_1\right)\lesssim2.3\times 10^{4}$
is always true if the lower threshold exists, $\alpha_1>-\alpha_1^{cr}=-4/27$.
The second condition $2.3\times 10^{5}\lesssim x_1^u\left(\alpha_1\right)$
is fulfilled if $\alpha_1\gtrsim-1.9\times 10^{-11}$. These two necessary
conditions can be translated into a constraint for $\xi_1$ using
the definition for $\alpha_n$, Eq.~(\ref{alpha_def}), and
$k_0\simeq4.4\times10^{14}\,$eV :
\begin{equation}\label{constr1}
\alpha_1\equiv\kappa_1\frac{k_0}{4\omega_b}
\left(\frac{k_0}{M_{\rm pl}}\right)\gtrsim-1.9\times 10^{-11}\,;
\quad \kappa_1\gtrsim-2.4\times 10^{-15}\,.
\end{equation}
For $n=1$ effective field theory implies LI violating terms in the dispersion
relation of equal absolute value and opposite sign for left and
right polarized photons~\cite{Myers:2003fd}.
Therefore, in order to avoid photon fractions in cosmic rays
$\gtrsim5$ times higher than observed above
$\sim10^{19}\,$eV, pair production has to be allowed for 
both polarizations, and thus for both signs in the dispersion relation. 
Thus the constraint obtained for $\kappa_1\simeq\xi_1<0$ is valid also 
for positive LI violating terms: $\left|\xi_1\right| \lesssim 2.4\times 10^{-15}$.

{\it Constraints on Lorentz invariance breaking to second order in the Planck mass.}
In this case $n=2$ and the first condition,
$x_2^l\left(\alpha_2\right)\lesssim2.3\times 10^{4}$
is always true if the lower threshold exists, $\alpha_2>-\alpha_2^{cr}=-27/256$.
The second condition $2.3\times 10^{5}\lesssim x_2^u\left(\alpha_2\right)$
is fulfilled if $\alpha_2\gtrsim-8.2\times 10^{-17}$. These two necessary conditions 
then lead to the following constraint for $\xi_2$
\begin{equation}\label{constr2}
\alpha_2\equiv\kappa_2\frac{k_0}{4\omega_b}
\left(\frac{k_0}{M_{\rm pl}}\right)^2\gtrsim-8.2\times 10^{-17}\,;
\quad \kappa_2\gtrsim-2.4\times 10^{-7}\,.
\end{equation}
For interactions with the URB, $k_0\simeq 6\times10^{19}\,$eV, we obtain
the constraint assuming the existence of at least one solution 
with $x_n^l(\alpha_n)\lesssim 2$. This eventually leads to the conditions
$\left|\kappa_1\right|\lesssim7.2\times 10^{-21}$ at first order, and
$-\kappa_2\lesssim8.5\times 10^{-13}$ at
second order. These are several orders of magnitudes more restrictive than
the constraints Eqs.~(\ref{constr1}) and~(\ref{constr2}) obtained in the CMB case. 
Therefore, if the constraints from interactions with the CMB are violated,
there would also be no interaction with the URB and so no pair production 
on any relevant background. Thus the constraint from pair production with the 
CMB is not modified by the presence of the URB.

{\it Discussion and Conclusions.}
To our knowledge, only  LI breaking suppressed to first order in the Planck
mass have so far been ruled out in the electromagnetic 
sector~\cite{Jacobson:2005bg,Liberati:2007vx,Maccione:2007yc}.
In terms of the dimensionless parameters $\xi_n$, the best
upper limit was $\left|\xi_1\right|\lesssim2\times 10^{-7}$ \cite{Fan:2007zb},
based on frequency dependent rotation of linear polarization (vacuum 
birefringence) of optical/UV photons of the afterglow from distant 
$\gamma-$ray bursts. A former, more stringent constraint,
$\left|\xi_1\right|\lesssim2\times 10^{-15}$ \cite{Jacobson:2003bn} was based on
polarization of MeV $\gamma-$rays which could not be
confirmed~\cite{Liberati:2007vx}.

Constraints based on modified reaction thresholds were so far
obtained from observations of multi-TeV $\gamma-$rays
from blazars at distances $\gtrsim100\,$Mpc, over which
such photons are expected to produce pairs on the infrared background.
However, given that involved photon energies are much smaller
than $10^{19}\,$eV, resulting constraints are of the order 
$\left|\xi_1\right|\lesssim1$~\cite{Stecker:2003pw}, much weaker than
our constraints $\left|\xi_1\right| \lesssim 2.4\times 10^{-15}$ and
$-\xi_2 \lesssim 2.4\times 10^{-7}$. Our new constraints
suggest that LI breaking suppressed up to second order
in the Planck scale are unlikely to be phenomenologically viable
for photons. Although similar constraints have been obtained in
an independent approach based on the absence of vacuum \v{C}erenkov
radiation of ultra-high energy protons~\cite{Gagnon:2004xh,Bernadotte:2006ya},
such constraints depend on the somewhat uncertain partonic structure
of these protons.

It is interesting
to note that the detection of a photon of $10^{19}$ eV would put
strong constraints on any positive LI breaking term in the dispersion
relation, $\kappa_1<10^{-17}$ for $n=1$ and $\kappa_2<10^{-8}$ for $n=2$,
in order to avoid photon decay.

Our constraints Eqs.~(\ref{constr1}) and~(\ref{constr2}) hold for the linear
combinations of LI breaking terms for photons, electrons and positrons
defined in Eq.~(\ref{alpha_def}). They translate directly
into constraints on the photon terms $\xi_n$ if the LI breaking terms for
electrons and positrons are significantly smaller than the ones
for photons. This is typically the case if the only
LI breaking terms for electrons/positrons are induced by their photon
content~\cite{Gagnon:2004xh}.

Note that in supersymmetric QED, corrections to the dispersion relation
of a particle of mass $m$ are of the form $\xi_n m^2(k/M_{\rm Pl})^n$ and
are thus negligible in astrophysical contexts~\cite{Groot Nibbelink:2004za}.
Therefore, our constraints only apply to the non-supersymmetric case.

{\it Acknowledgements.}
We are grateful to Theodore A. Jacobson and David Mattingly for valuable
comments on this project.
MG thanks APC for the hospitality during the developments of this work.


\begin{thebibliography}{99}

\bibitem{Coleman:1998ti}
  S.~R.~Coleman and S.~L.~Glashow,
  Phys.\ Rev.\  D {\bf 59}, 116008 (1999)
  [arXiv:hep-ph/9812418].

\bibitem{gzk}
K.~Greisen,
Phys.\ Rev.\ Lett.\  {\bf 16}, 748 (1966);
G.~T.~Zatsepin and V.~A.~Kuzmin,
JETP Lett.\  {\bf 4}, 78 (1966)
[Pisma Zh.\ Eksp.\ Teor.\ Fiz.\  {\bf 4}, 114 (1966)].

\bibitem{Aloisio:2000cm}
  R.~Aloisio, P.~Blasi, P.~L.~Ghia and A.~F.~Grillo,
  Phys.\ Rev.\  D {\bf 62}, 053010 (2000)
  [arXiv:astro-ph/0001258].
  
\bibitem{Jacobson:2002hd}
  T.~Jacobson, S.~Liberati and D.~Mattingly,
  Phys.\ Rev.\  D {\bf 67}, 124011 (2003)
  [arXiv:hep-ph/0209264].

\bibitem{Abraham:2006ar}
  J.~Abraham {\it et al.}  [Pierre Auger Collaboration],
  Astropart.\ Phys.\  {\bf 27}, 155 (2007)
  [arXiv:astro-ph/0606619].

\bibitem{Shinozaki:2002ve}
  K.~Shinozaki {\it et al.},
  Astrophys.\ J.\  {\bf 571}, L117 (2002).
  
\bibitem{Risse:2005jr}
  M.~Risse {\it et al.},
  Phys.\ Rev.\ Lett.\  {\bf 95}, 171102 (2005)
  [arXiv:astro-ph/0502418].

\bibitem{Rubtsov:2006tt}
  G.~I.~Rubtsov {\it et al.},
  Phys.\ Rev.\ D {\bf 73}, 063009 (2006)
  [arXiv:astro-ph/0601449].

\bibitem{Glushkov:2007ss}
  A.~V.~Glushkov, D.~S.~Gorbunov, I.~T.~Makarov, M.~I.~Pravdin, G.~I.~Rubtsov, I.~E.~Sleptsov and S.~V.~Troitsky,
  arXiv:astro-ph/0701245.

\bibitem{Healy:2007ef}
  M.~D.~Healy for the Pierre~Auger~Collaboration,
  arXiv:0710.0025 [astro-ph].

\bibitem{Risse:2007sd}
  M.~Risse and P.~Homola,
  Mod.\ Phys.\ Lett.\  A {\bf 22}, 749 (2007)
  [arXiv:astro-ph/0702632].

\bibitem{Sigl:2007ea}
  G.~Sigl,
  Phys.\ Rev.\  D {\bf 75}, 103001 (2007)
  [arXiv:astro-ph/0703403].

\bibitem{Gelmini:2007jy}
  G.~B.~Gelmini, O.~Kalashev and D.~V.~Semikoz,
  arXiv:0706.2181 [astro-ph].

\bibitem{crpropa} see {\sf http://apcauger.in2p3.fr//CRPropa}.

\bibitem{Armengaud:2006fx}
  E.~Armengaud, G.~Sigl, T.~Beau and F.~Miniati,
  arXiv:astro-ph/0603675.

\bibitem{Shinozaki:2006kk}
  K.~Shinozaki  [AGASA Collaboration],
  Nucl.\ Phys.\ Proc.\ Suppl.\  {\bf 151}, 3 (2006);
  see also {\sf http~://www-akeno.icrr.u-tokyo.ac.jp/AGASA/}.

\bibitem{Abbasi:2002ta}
  R.~U.~Abbasi {\it et al.}  [High Resolution Fly's Eye Collaboration],
  Phys.\ Rev.\ Lett.\  {\bf 92}, 151101 (2004)
  [arXiv:astro-ph/0208243].

\bibitem{radio-obs} T.~A.~Clark, L.~W.~Brown, J.~K.~Alexander, Nature {\bf 228},
847 (1970).

\bibitem{Gagnon:2004xh}
  O.~Gagnon and G.~D.~Moore,
  Phys.\ Rev.\  D {\bf 70}, 065002 (2004)
  [arXiv:hep-ph/0404196].

\bibitem{Mattingly:2002ba}
  D.~Mattingly, T.~Jacobson and S.~Liberati,
  Phys.\ Rev.\  D {\bf 67}, 124012 (2003)
  [arXiv:hep-ph/0211466].

\bibitem{Myers:2003fd}
  R.~C.~Myers and M.~Pospelov,
  Phys.\ Rev.\ Lett.\  {\bf 90} (2003) 211601
  [arXiv:hep-ph/0301124].

\bibitem{Jacobson:2005bg}
  T.~Jacobson, S.~Liberati and D.~Mattingly,
  Annals Phys.\  {\bf 321}, 150 (2006)
  [arXiv:astro-ph/0505267].

\bibitem{Liberati:2007vx}
  S.~Liberati,
  PoS {\bf P2GC}, 018 (2007)
  [arXiv:0706.0142 [gr-qc]].

\bibitem{Maccione:2007yc}
  L.~Maccione, S.~Liberati, A.~Celotti and J.~G.~Kirk,
  arXiv:0707.2673 [astro-ph].

\bibitem{Fan:2007zb}
  Y.~Z.~Fan, D.~M.~Wei and D.~Xu,
  Mon.\ Not.\ Roy.\ Astron.\ Soc.\  {\bf 376}, 1857 (2006)
  [arXiv:astro-ph/0702006].

\bibitem{Jacobson:2003bn}
  T.~A.~Jacobson, S.~Liberati, D.~Mattingly and F.~W.~Stecker,
  Phys.\ Rev.\ Lett.\  {\bf 93}, 021101 (2004)
  [arXiv:astro-ph/0309681].

\bibitem{Stecker:2003pw}
  F.~W.~Stecker,
  Astropart.\ Phys.\  {\bf 20}, 85 (2003)
  [arXiv:astro-ph/0308214].

\bibitem{Bernadotte:2006ya}
  S.~Bernadotte and F.~R.~Klinkhamer,
  Phys.\ Rev.\  D {\bf 75}, 024028 (2007)
  [arXiv:hep-ph/0610216].

\bibitem{Groot Nibbelink:2004za}
  S.~Groot Nibbelink and M.~Pospelov,
  Phys.\ Rev.\ Lett.\  {\bf 94}, 081601 (2005)
  [arXiv:hep-ph/0404271].

\end{thebibliography}
\end{document}